\documentclass[12pt,preprint]{aastex}
\usepackage{amsmath,amssymb}
\usepackage{graphicx}
\usepackage{float}
\usepackage{multirow}

\slugcomment{KSUPT-13/1 \hspace{0.5truecm} January 2013}

\begin{document}

\title{Hubble parameter measurement constraints on the cosmological 
deceleration-acceleration transition redshift}

\author{Omer Farooq and Bharat Ratra}

\affil{Department of Physics, Kansas State University, 
                 116 Cardwell Hall, Manhattan, KS 66506, USA} 
\email{omer@phys.ksu.edu, ratra@phys.ksu.edu}

\begin{abstract}

We compile a list of 28 independent measurements of the Hubble parameter 
between redshifts $0.07 \leq z \leq 2.3$ and use this to place constraints 
on model parameters of constant and time-evolving dark energy cosmologies. 
These $H(z)$ measurements by themselves require a currently accelerating
cosmological expansion at about, or better than, 3 $\sigma$ confidence.
The mean and standard deviation of the 6 best-fit model 
deceleration-acceleration
transition redshifts (for the 3 cosmological models and 2 Hubble
constant priors we consider) is $z_{\rm da} = 0.74 \pm 0.05$, in good
agreement with the recent \citet{busca12} determination of 
$z_{\rm da} = 0.82 \pm 0.08$ based on 11 $H(z)$ measurements between redshifts 
$0.2 \leq z \leq 2.3$, almost entirely from BAO-like data.

\end{abstract}

\maketitle
\section{Introduction}
\label{intro}

In the standard picture of cosmology, dark energy powers the current 
accelerating cosmological expansion but played a less significant role 
in the past when nonrelativistic (cold dark and baryonic) matter 
dominated and powered the then decelerating cosmological 
expansion.\footnote{
For reviews of dark energy see \cite{Bolotin2011}, 
\cite{Martin2012}, and references therein. The observed accelerating 
cosmological expansion has also be interpreted as indicating the
need to modify general relativity. In this paper we assume that general 
relativity provides an adequate description of gravitation on 
cosmological length scales. For reviews of modified gravity see 
\cite{Bolotin2011}, \cite{Capozziello2011}, and references therein.}
It is of some interest to determine the redshift of the 
deceleration-acceleration transition predicted to exist in dark energy 
cosmological models. There have been a number of attempts to do
so, see, e.g., \citet{Lu2011a}, \cite{Giostri2012}, \cite{Lima2012},
and references therein. However, until very recently, this has not been 
possible because there has not been much high-quality data at high enough 
redshift (i.e., for $z$ above the transition redshift in standard dark 
energy cosmological models).

The recent \cite{busca12} detection of the baryon acoustic oscillation 
(BAO) peak at $z = 2.3$ in the Ly$\alpha$ forest has dramatically 
changed the situation by allowing for a high precision measurement of the 
Hubble parameter $H(z)$ at $z = 2.3$, well in the matter dominated epoch
of the standard dark energy cosmological model. \citet{busca12} use 
this and 10 other $H(z)$ measurements, largely based on BAO-like data,
and the \cite{Riess2011} HST determination of the Hubble constant, in the 
context of the standard $\Lambda$CDM cosmological model, to estimate a 
deceleration-acceleration transition redshift of $z_{\rm da} = 0.82 \pm 0.08$.

In this paper we extend the analysis of \cite{busca12}. We first compile
a list of 28 independent $H(z)$ measurements.\footnote{
It appears that some of the measurements listed in Table 2 of 
\cite{busca12} might not be independent. For instance, the \cite{Chuang2012a}
and the \cite{Xu2012b} determinations of $H(z = 0.35)$ listed in the table 
are both based on the use of Sloan Digital Sky Survey Data Release 7 
measurements of luminous red galaxies.}
We then use these 28 measurements to constrain cosmological parameters 
in 3 different dark energy models and establish that the models are
a good fit to the data and that the data provide tight constraints on the 
model parameters. Finally we use the models to estimate the redshift of the 
deceleration-acceleration transition. \cite{busca12} have one measurement 
(of 11)
above their estimated $z_{\rm da} = 0.82$, while we have 9 of 28 above this
(and 10 of 28 above our estimated redshift $z_{\rm da} = 0.74$). 
Granted, the \cite{busca12} $z$ = 2.3 measurement carries great weight because 
of the small, 3.6\%, uncertainity, but 9 of our 10 high redshift
measurements, from \cite{simon05}, \cite{Stern2010}, and \cite{moresco12}, 
include 3 11\%, 13\%, and 14\% measurements from \cite{moresco12} and
3 10\% measurements from \cite{simon05}, all 6 of which carry significant 
weight. 

Dark energy, most simply thought of as a negative pressure substance,  
dominates the current cosmological energy budget. In this paper we consider 
3 dark energy models.

The first one is the ``standard" spatially-flat $\Lambda$CDM cosmological
model \citep{peebles84}. In this model a little over $70\%$ of the current 
energy budget is dark energy (Einstein's cosmological constant $\Lambda$),
non-relativistic cold dark matter (CDM) being the next largest contributer 
(a little over $20\%$), followed by non-relativistic baryonic matter 
(about $5\%$). In the $\Lambda$CDM model the dark energy density is 
constant in time and does not vary in space. $\Lambda$CDM has a 
number of well-known puzzling features 
\citep[see, e.g.,][]{Peebles&Ratra2003}. 

These puzzles could be eased if the dark energy density is a slowly 
decreasing function of time \citep[][]{Ratra&Peebles1988}.\footnote{ 
For recent discussions of time-varying dark energy models, see
\cite{Gu2012}, \cite{Basilakos2012}, \cite{Xu2012a}, \cite{Guendelman2012}, 
and references therein.} 
In this paper we consider a slowly-evolving dark energy scalar field 
model as well as a time-varying dark energy parameterization.

In $\Lambda$CDM, time-independent dark energy density is modeled as 
a spatially homogeneous fluid with equation of state $p_{\rm \Lambda} 
= -\rho_{\rm \Lambda}$ where $p_{\rm \Lambda}$ and $\rho_{\rm \Lambda}$ 
are the fluid pressure and energy density. Much use has been made of a 
parametrization of slowly-decreasing dark energy density known as XCDM
where dark energy is modeled as a spatially homogeneous fluid with 
equation of state $p_{\rm X}=w_{\rm X}\rho_{\rm X}$. The equation of 
state parameter $w_{\rm X}<-1/3$ is independent of time and $p_{\rm X}$ 
and $\rho_{\rm X}$ are the  pressure and energy density of the $X$-fluid. 
When $w_{\rm X}=-1$ the XCDM parameterization reduces 
to the complete and consistent $\Lambda$CDM model. For any other 
value of $w_{\rm X}<-1/3$ the XCDM parameterization is incomplete as it 
cannot describe spatial inhomogeneities \citep[see, e.g.][]{ratra91, 
podariu2000}. For computational simplicity, in the 
XCDM case we assume a spatially-flat cosmological model.  
The $\phi$CDM model is the simplest, consistent and complete model of
slowly-decreasing dark energy density \citep{Ratra&Peebles1988}. 
Here dark energy is modeled as a scalar field, $\phi$, with a gradually 
decreasing (in $\phi$) potential energy density $V(\phi)$. In this paper
we assume an inverse power-law potential energy 
density $V(\phi) \propto \phi^{-\alpha}$, where $\alpha$ is a nonnegative 
constant \citep{Peebles&Ratra1988}. When $\alpha = 0$ the
$\phi$CDM model reduces to the corresponding $\Lambda$CDM case. 
For computational simplicity, we again only consider the spatially-flat
cosmological case for $\phi$CDM. 

\begin{table}
\begin{center}
\begin{tabular}{cccc}
\hline\hline
~~$z$ & ~~$H(z)$ &~~~~~~~ $\sigma_{H}$ &~~ Reference\\
~~~~~    & (km s$^{-1}$ Mpc $^{-1}$) &~~~~~~~ (km s$^{-1}$ Mpc $^{-1}$)& \\
\tableline
0.070&~~	69&~~~~~~~	19.6&~~ 5\\
0.100&~~	69&~~~~~~~	12&~~	1\\
0.120&~~	68.6&~~~~~~~	26.2&~~	5\\
0.170&~~	83&~~~~~~~	8&~~	1\\
0.179&~~	75&~~~~~~~	4&~~	3\\
0.199&~~	75&~~~~~~~	5&~~	3\\
0.200&~~	72.9&~~~~~~~	29.6&~~	5\\
0.270&~~	77&~~~~~~~	14&~~	1\\
0.280&~~	88.8&~~~~~~~	36.6&~~	5\\
0.350&~~	76.3&~~~~~~~	5.6&~~	7\\
0.352&~~	83&~~~~~~~	14&~~	3\\
0.400&~~	95&~~~~~~~	17&~~	1\\
0.440&~~	82.6&~~~~~~~	7.8&~~	6\\
0.480&~~	97&~~~~~~~	62&~~	2\\
0.593&~~	104&~~~~~~~	13&~~	3\\
0.600&~~	87.9&~~~~~~~	6.1&~~	6\\
0.680&~~	92&~~~~~~~	8&~~	3\\
0.730&~~	97.3&~~~~~~~	7.0&~~	6\\
0.781&~~	105&~~~~~~~	12&~~	3\\
0.875&~~	125&~~~~~~~	17&~~	3\\
0.880&~~	90&~~~~~~~	40&~~	2\\
0.900&~~	117&~~~~~~~	23&~~	1\\
1.037&~~	154&~~~~~~~	20&~~	3\\
1.300&~~	168&~~~~~~~	17&~~	1\\
1.430&~~	177&~~~~~~~	18&~~	1\\
1.530&~~	140&~~~~~~~	14&~~	1\\
1.750&~~	202&~~~~~~~	40&~~	1\\
2.300&~~	224&~~~~~~~	8&~~	4\\

\hline\hline
\end{tabular}
\end{center}
\caption{Hubble parameter versus redshift data. Last column reference numbers:
1.\ \cite{simon05}, 2.\ \cite{Stern2010}, 3.\ \cite{moresco12}, 4.\ 
\cite{busca12}, 5.\ \cite{Zhang2012}, 6.\ \cite{blake12}, 7.\ 
\cite{Chuang2012b}.
}\label{tab:Hz}
\end{table}

Many different data sets have been used to derive constraints on the
3 cosmological models we consider here.\footnote{
See, e.g., \citet{chae04}, \citet{Samushia&Ratra2008}, \citet{Lu2011b},
\citet{dantas11}, \citet{Cao2012}, \citet{Chen2012}, \citet{Jackson2012},
\citet{campanelli11}, \citet{Poitras2012}, and 
references therein.}
Of interest to us here are measurements of the Hubble parameter as a 
function of redshift \citep[e.g.,][]{Jimenezetal2003, Samushia&Ratra2006,
samushia07, Sen&Scherrer2008, Chen2011b, Duan2011, Aviles2012, seikel12}. 
Table 1 lists 28 $H(z)$ measurements. We only include
independent measurements of $H(z)$, listing only the most recent 
result from analyses of a given data set. The values in Table 1 have been 
determined using a number of different techniques; for details see 
the papers listed in the table caption. Table 1 is the largest set of 
independent $H(z)$ measurements considered to date. 

We first use these data to derive constraints on cosmological parameters
of the 3 models described above. The constraints derived here are compatible 
with cosmological parameter constraints determined by other techniques. 
These constraints are more restrictive than those derived by \cite{Farooq2012b}
using the previous largest set of $H(z)$ measurements, as well as those
derived from the recent SNIa data compilation of \cite{suzuki12}. 
The $H(z)$ data considered here require accelerated cosmological expansion 
at the current epoch at about or more than 3 $\sigma$ confidence.

Our paper is organized as follows. In the next section we
present constraints from the $H(z)$ data on cosmological 
parameters of the 3 models we consider, establish that the 
3 models are very consistent with the $H(z)$ data, and use the 
models to estimate the redshift of the cosmological 
deceleration-acceleration transition. We conclude in 
Sec.\ {\ref{summary}}.

\section{Constraints from the $H(z)$ data}
\label{HzData}

\begin{figure}[t]
\centering
  \includegraphics[width=100mm]{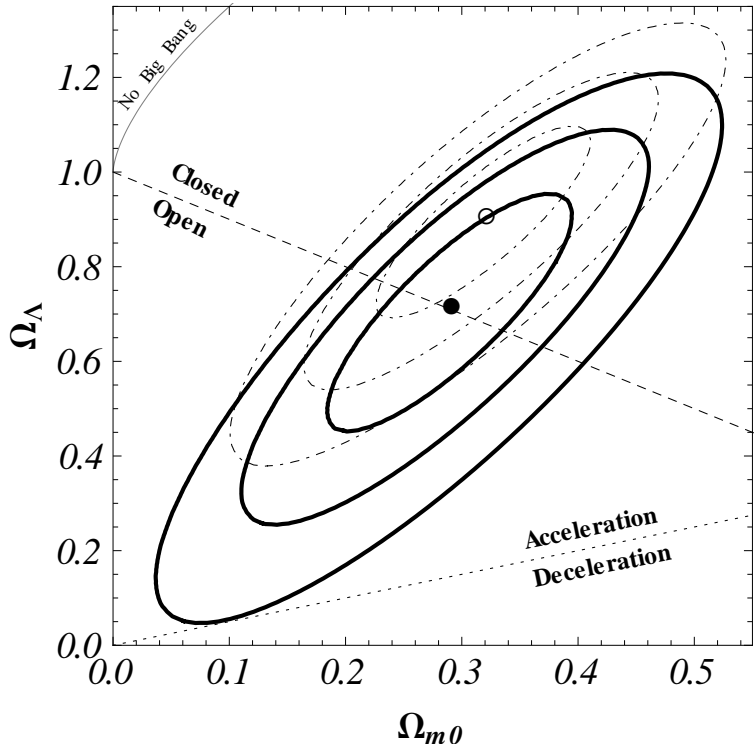}
\caption{
Solid [dot-dashed] lines show 1, 2, and 3 $\sigma$ constraint contours for 
the $\Lambda$CDM model from the $H(z)$ data given in Table \ref{tab:Hz} for 
the prior $\bar{H}_0 \pm \sigma_{H_0} = 68 \pm 2.8$ km s$^{-1}$ Mpc$^{-1}$
[$\bar{H}_0 \pm \sigma_{H_0} = 73.8 \pm 2.4$ km s$^{-1}$ Mpc$^{-1}$].
The filled [empty] circle best-fit point is at 
$(\Omega_{m0},\Omega_{\Lambda})=(0.29,0.72)$ 
[$(0.32,0.91)$] with $\chi^2_{\rm min}=18.24$ [$19.30$].
The dashed diagonal line corresponds to spatially-flat models, the 
dotted line demarcates zero-acceleration models, and the area 
in the upper left-hand corner is the region for which there is no 
big bang. The 2 $\sigma$ intervals from the one-dimensional marginalized 
probability distributions are $0.15\leqslant \Omega_{m0}\leqslant 0.42$, 
$0.35\leqslant \Omega_{\Lambda}\leqslant 1.02$ 
[$0.20\leqslant \Omega_{m0}\leqslant 0.44$, 
$0.62\leqslant \Omega_{\Lambda}\leqslant 1.14$].
} \label{fig:LCDM_Hz}
\end{figure}

Following \cite{Farooq2012}, we use the 
28 independent $H(z)$ data points listed in Table \ref{tab:Hz} 
to constrain cosmological model parameters. 
The observational data consist of measurements of the 
Hubble parameter $H_{\rm obs}(z_i)$ at redshifts $z_i$, with the 
corresponding one standard deviation uncertainties $\sigma_i$.
To constrain cosmological parameters $\textbf{p}$ of the models of 
interest we build the posterior likelihood function 
$\mathcal{L}_{H}(\textbf{p})$ that depends only on the 
$\textbf{p}$ by integrating the product of exp$(-\chi_H^2 /2)$ and 
the $H_0$ prior likelihood function 
exp$[-(H_0-\bar H_0)^2/(2\sigma^2_{H_0})]$, as in Eq.\ 18 of 
\cite{Farooq2012}. We marginalize over the nuisance parameter $H_0$ 
using two different Gaussian priors with $\bar{H_0}\pm\sigma_{H_0}$=
68 $\pm$ 2.8 km s$^{-1}$ Mpc$^{-1}$ \citep{Chen2003, Chen2011a}
and with $\bar{H_0}\pm\sigma_{H_0}$ = 73.8 
$\pm$ 2.4 km s$^{-1}$ Mpc$^{-1}$ \citep{Riess2011}. As discussed there, 
the Hubble constant measurement uncertainty can significantly affect 
cosmological parameter estimation \citep[for a recent example see, 
e.g.,][]{calabrese12}. We determine the parameter values that maximize 
the likelihood function and find 1, 2, and 3 $\sigma$ constraint 
contours by integrating the likelihood function, starting from the 
maximum and including 68.27 \%, 95.45 \%, and 99.73 \% of the probability.

\begin{figure}[t]
\centering
  \includegraphics[width=100mm]{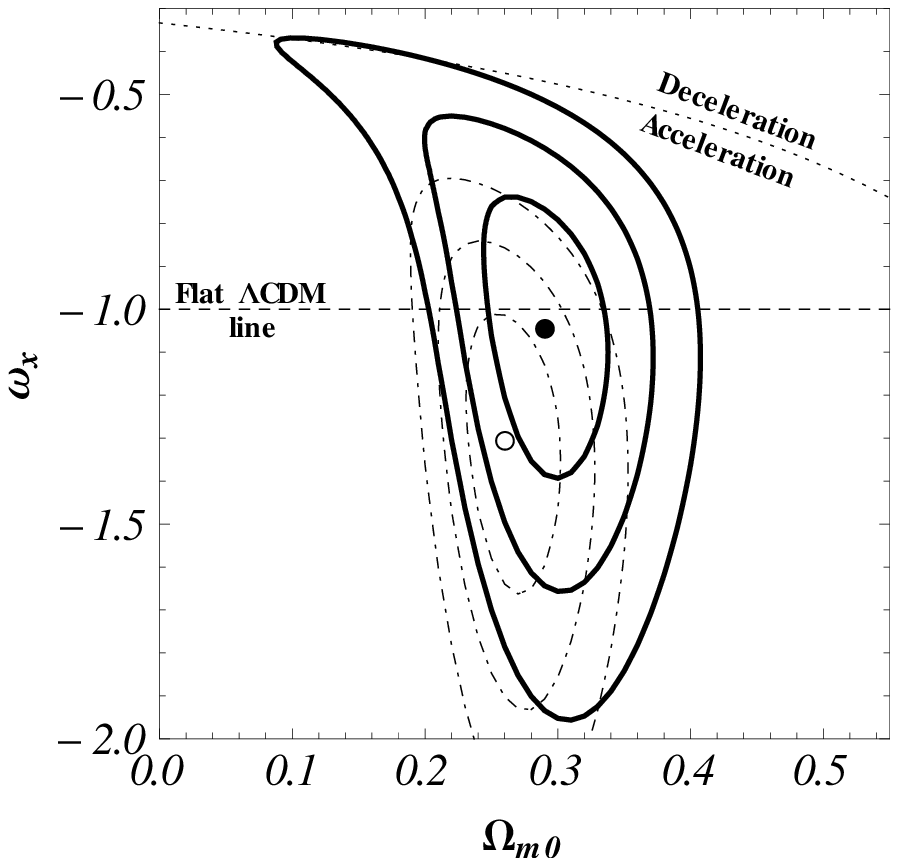}
\caption{
Solid [dot-dashed] lines show 1, 2, and 3 $\sigma$ constraint contours for 
the XCDM parametrization from the $H(z)$ data given in Table \ref{tab:Hz} 
for the prior $\bar{H}_0 \pm \sigma_{H_0} = 68 \pm 2.8$ km s$^{-1}$ Mpc$^{-1}$
[$\bar{H}_0 \pm \sigma_{H_0} = 73.8 \pm 2.4$ km s$^{-1}$ Mpc$^{-1}$].
The filled [empty] circle is the best-fit point at
$(\Omega_{m0},\omega_{X})=(0.29,-1.04)$  [$(0.26,-1.30)$] 
with $\chi^2_{\rm min}=18.18$ [$18.15$].
The dashed horizontal line at $\omega_{\rm X} = -1$ corresponds to 
spatially-flat $\Lambda$CDM models and the curved dotted line demarcates 
zero-acceleration models. The 2 $\sigma$ intervals from the one-dimensional 
marginalized probability distributions are 
$0.23\leqslant \Omega_{m0}\leqslant 0.35$, 
$-1.51\leqslant \omega_{X}\leqslant -0.64$ 
[$0.22\leqslant \Omega_{m0}\leqslant 0.31$, 
$-1.78\leqslant \omega_{X}\leqslant -0.92$].
} \label{fig:XCDM_Hz}
\end{figure}

\begin{figure}[t]
\centering
  \includegraphics[width=100mm]{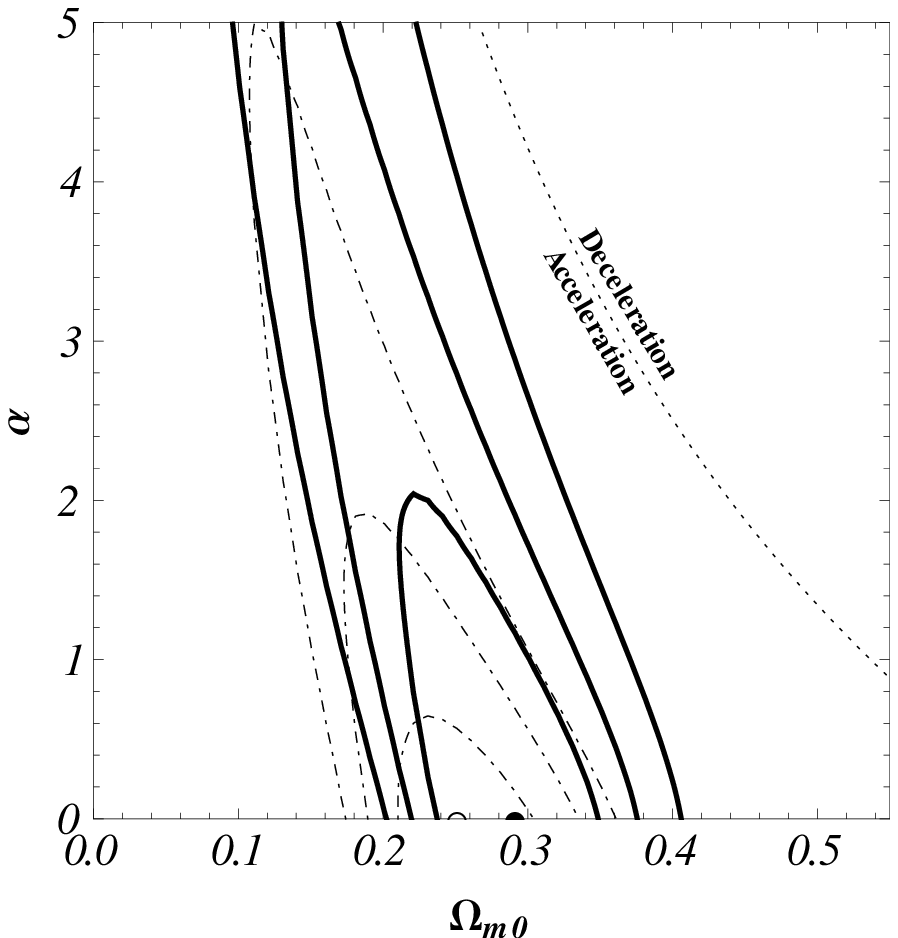}
\caption{
Solid [dot-dashed] lines show 1, 2, and 3 $\sigma$ constraint contours for 
the $\phi$CDM model from the $H(z)$ data given in Table \ref{tab:Hz} for 
the prior $\bar{H}_0 \pm \sigma_{H_0} = 68 \pm 2.8$ km s$^{-1}$ Mpc$^{-1}$
[$\bar{H}_0 \pm \sigma_{H_0} = 73.8 \pm 2.4$ km s$^{-1}$ Mpc$^{-1}$].
The filled [empty] circle best-fit point is at $(\Omega_{m0},\alpha)=(0.29,0)$ 
[$(0.25,0)$] with $\chi^2_{\rm min}=18.24$ [$20.64$].
The horizontal axis at $\alpha = 0$ corresponds to spatially-flat 
$\Lambda$CDM models and the curved dotted line demarcates 
zero-acceleration models. The 2 $\sigma$ intervals from the one-dimensional 
marginalized probability distributions are 
$0.17\leqslant \Omega_{m0}\leqslant 0.34$, 
$\alpha \leqslant 2.2$ [$0.16\leqslant \Omega_{m0}\leqslant 0.34$, 
$\alpha \leqslant 0.7$].
} \label{fig:phiCDM_Hz}
\end{figure}

Figures \ref{fig:LCDM_Hz}---\ref{fig:phiCDM_Hz} show the constraints
from the $H(z)$ data for the three dark energy models we consider, and
for the two different $H_0$ priors. In all 6 cases the $H(z)$ data of
Table 1 require accelerated cosmological expansion at the current epoch,
at, or better than, 3 $\sigma$ confidence. The previous largest 
$H(z)$ data set used, that in \cite{Farooq2012b}, required this
accelerated expansion at, or better than, 2 $\sigma$ confidence.
Comparing Figs.\ 1---3 here to Figs.\ 1---3 of \cite{Farooq2012b}, we
see that in the XCDM and $\phi$CDM cases the $H(z)$ data we use in
this paper significantly tightens up the constraints on 
$w_{\rm X}$ and $\alpha$, but does not much affect the $\Omega_{m0}$
constraints. However, in the $\Lambda$CDM case the $H(z)$ data used here
tightens up constraints on both $\Omega_\Lambda$ and $\Omega_{m0}$.

\begin{figure}[t]
\centering
  \includegraphics[width=140mm]{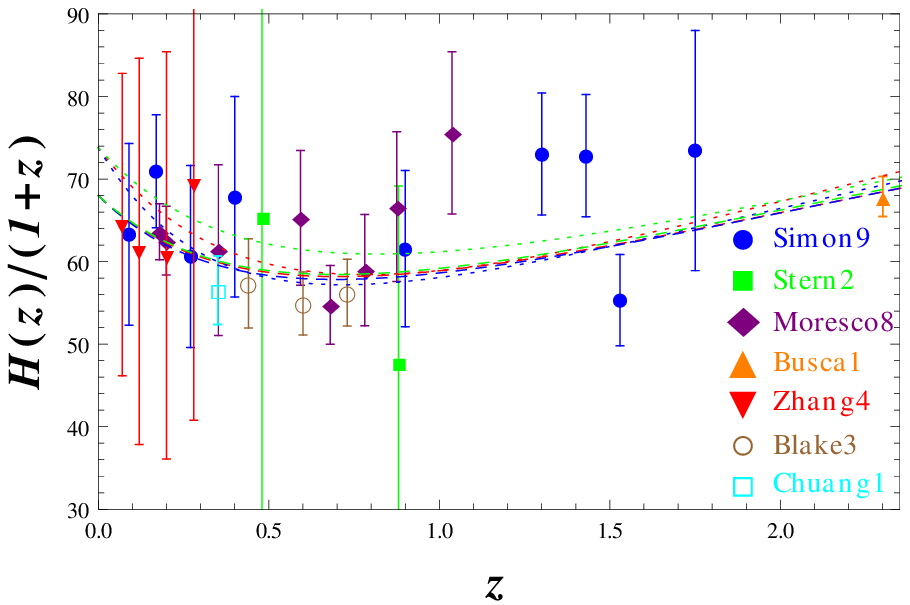}
\caption{
$H(z)/(1+z)$ data (28 points) and model predictions (lines for 6 
best-fit models) as a function of redshift. The dashed [dotted] lines 
are for the prior $\bar{H}_0 \pm \sigma_{H_0} = 
68 \pm 2.8$ km s$^{-1}$ Mpc$^{-1}$ 
[$\bar{H}_0 \pm \sigma_{H_0} = 73.8 \pm 2.4$ km s$^{-1}$ Mpc$^{-1}$], with
red, blue, and green lines corresponding to the $\Lambda$CDM, XCDM, and 
$\phi$CDM cases.
} \label{fig:Hz+modelfit}
\end{figure}

As indicated by the $\chi^2_{\rm min}$ values listed in the captions of 
Figs.\ 1---3, all 6 best-fit models are very consistent with the $H(z)$
data listed in Table 1. It is straightforward to compute the cosmological
deceleration-acceleration transition redshift in these cases. They are
0.706 [0.785], 0.695 [0.718], and 0.698 [0.817] for the $\Lambda$CDM, 
XCDM, and $\phi$CDM  models with prior 
$\bar{H}_0 \pm \sigma_{H_0} = 68 \pm 2.8$ km s$^{-1}$ Mpc$^{-1}$ 
[$\bar{H}_0 \pm \sigma_{H_0} = 73.8 \pm 2.4$ km s$^{-1}$ Mpc$^{-1}$].
The mean and standard deviation give $z_{\rm da} = 0.74 \pm 0.05$, in good
agreement with the recent \citet{busca12} determination of 
$z_{\rm da} = 0.82 \pm 0.08$ based on less data, possibly not all 
independent. Figure \ref{fig:Hz+modelfit} shows $H(z)/(1+z)$ data from 
Table \ref{tab:Hz} and the 6 best-fit model predictions as a function of 
redshift. The deceleration-acceleration transition is not impossible
to discern in the data.

\section{Conclusion}
\label{summary}

In summary, we have extended the analysis of \cite{busca12} to a larger
independent set of 28 $H(z)$ measurements and determined the 
cosmological deceleration-acceleration transition redshift 
$z_{\rm da} = 0.74 \pm 0.05$. These $H(z)$ data are well-described by all
6 best-fit models, and provide tight constraints on the model
parameters. The $H(z)$ data require accelerated cosmological expansion 
at the current epoch, and are consistent with the decelerated 
cosmological expansion at earlier times predicted and required in standard 
dark energy models. While the standard spatially-flat 
$\Lambda$CDM model is very consistent with the $H(z)$ data, current
$H(z)$ data are not able to rule out slowly evolving dark energy.
More, and better quality, data are needed to better discriminate
between constant and slowly-evolving dark energy density; these
data are likely to soon be in hand.

\acknowledgments

We thank Mikhail Makouski and Data Mania for useful discussions and 
helpful advice. This work was supported in part by DOE grant 
DEFG03-99EP41093 and NSF grant AST-1109275.


\end{document}